\documentclass{./ActaPhysicaPolonicaB}

\usepackage{graphicx}
\usepackage{hyperref}
\usepackage[utf8]{inputenc}
\usepackage{amsmath}
\usepackage{xspace}
\usepackage{color}
\usepackage{lineno}

\newcommand{\pp}{\mbox{pp}\xspace}
\newcommand{\PbPb}{\mbox{Pb--Pb}\xspace}
\newcommand{\pPb}{\mbox{p--Pb}\xspace}
\newcommand{\five}{$\sqrt{s}~=~5.02$~Te\kern-.1emV\xspace}
\newcommand{\fivenn}{$\sqrt{s_{\mathrm{NN}}}~=~5.02$~Te\kern-.1emV\xspace}
\newcommand{\RpPb}    {\ensuremath{R_{\rm pPb}}\xspace}
\newcommand{\RpPbjch}{\ensuremath{\RpPb^{\rm ch~jet}}\xspace}
\newcommand{\GeVc}{Ge\kern-.1emV/$c$\xspace}

\newcommand{\pTjch}{\ensuremath{p_{\mathrm{T}}^{\mathrm{ch~jet}}}\xspace}
\newcommand{\pT}{\ensuremath{p_{\mathrm T}}\xspace}
\newcommand{\pTj}{\ensuremath{p_{\mathrm{T,jet}}}\xspace}
\newcommand{\IAA}{\ensuremath{I_{\mathrm{AA}}(\pTjch)\xspace}}
\newcommand{\RL}{\ensuremath{R_{\mathrm{L}}}\xspace}
\newcommand{\pTjetch}{\ensuremath{p_\mathrm{T,jet}^\mathrm{ch}}\xspace}
\newcommand{\Ntrig}{\ensuremath{N_{\mathrm{trig}}}\xspace}
\newcommand{\TTSig}{\ensuremath{\mathrm{TT}_{\mathrm{Sig}}}\xspace}
\newcommand{\TTRef}{\ensuremath{\mathrm{TT}_{\mathrm{Ref}}}\xspace}
\newcommand{\dNjetdpTdphi}{\ensuremath{\frac{{\mathrm{d}}^{2}N_{\mathrm{jet}}}{\mathrm{d}\pTjetch\mathrm{d}\dphi}}\xspace}

\newcommand{\Drecoil}{\ensuremath{\Delta_\mathrm{recoil}}\xspace}
\newcommand{\DrecoilpTphi}{\ensuremath{\Delta_\mathrm{recoil}(\pT,\dphi)}\xspace}
\newcommand{\cRef}{\ensuremath{c_\mathrm{Ref}}\xspace}
\newcommand{\dphi}{\ensuremath{\Delta\varphi}\xspace}
\newcommand{\pTtrig}{\ensuremath{p_{\mathrm{T,trig}}}\xspace}

\begin{document}
\title{Jet and jet substructure: ALICE Results%
\thanks{Presented at ``Diffraction and Low-$x$ 2024'', Trabia (Palermo, Italy), September 8-14, 2024.}
}
\author{Haidar Mas'ud Alfanda\\
on behalf of the ALICE Collaboration
\address{Institute of Particle Physics and Key Laboratory of Quark and Lepton, Central China Normal University}
\\[3mm]
}
\maketitle
\begin{abstract}
Jets and their substructure in \pp collisions offer a unique opportunity to probe various aspects of quantum chromodynamics (QCD), ranging from perturbative QCD (pQCD) tests to studies of non-perturbative phenomena such as hadronization.
They also probe the transition between perturbative and non-perturbative regimes.
In heavy-ion collisions, jets serve as a novel tool to investigate the microscopic properties of the deconfined quark\textendash  gluon plasma (QGP).
Recently, significant progress has been made in developing jet substructure observables to explore these properties.
The ALICE experiment is particularly well-suited for jet measurements due to its high-precision tracking system, which is especially beneficial for detecting low transverse momentum jets.
This contribution will highlight recent ALICE measurements of inclusive and semi-inclusive jets, along with various jet substructure observables in \pp, \pPb, and \PbPb collisions.
The comparisons between data and predictions from Monte Carlo (MC) models as well as analytical calculations will be discussed.\end{abstract}

\section{Introduction}
High-energy heavy-ion collisions at the Large Hadron Collider (LHC) create a hot and dense form of matter called QGP.
The QGP created in the collision rapidly expands as a strongly-coupled liquid and cools down until a temperature near the phase transition at which the deconfined partons hadronize into ordinary color-neutral matter.
Scattering processes at very large momentum transfer $(Q^{2})$ between quarks and gluons of the colliding nucleons produce parton showers, which subsequently fragment into collimated sprays of hadrons called jets.
Jets play a pivotal role in the study and comprehension of QCD.
Moreover, jets are crucial probes for investigating the properties of the QGP.
In this proceeding, recent jet measurements are reported using the data collected with the ALICE detector during Run $2$. 
The excellent track reconstruction capabilities of the ALICE experiment were exploited to reconstruct jets using charged-particle tracks down to low transverse momentum~\cite{ALICE:2022wpn}.

\section{Jet energy-energy correlators in \pp collisions}
The jet energy-energy correlator (EEC) provides a powerful way to study the internal structure of jets and the transition from the perturbative to non-perturbative regimes of QCD~\cite{ALICE:2022wpn}.
The energy-energy correlation function, $\Sigma_{\text{EEC}}(\RL)$, is defined as
\begin{equation}
    \Sigma_{\text{EEC}}(\RL) = \frac{1}{ N_{\text{jet}}\cdot\Delta}\int_{\RL-\frac{1}{2}\Delta}^{\RL+\frac{1}{2}\Delta}\sum_{\text{jets}}\sum_{i, j}\frac{p_{\text{T}, i}p_{\text{T}, j}}{(\pTj)^{2}}\delta(\RL'-R_{\text{L}, ij})\text{d}\RL',
\end{equation}
\noindent
where $R_{\text{L}, ij} = \sqrt{(\varphi_{j} - \varphi_{i})^{2} + (\eta_{j} - \eta_{i})^{2}}$ is the relative separation between two particles $i$ and $j$ inside the jet, $\Delta$ is the angular bin width, and $N_{\text{jet}}$ is the total number of jets.
The scaling behavior of the EEC as a function of \RL reveals two distinct regions: a perturbative regime at large distances and a non-perturbative regime at small distances.
This separation provides insights into the dynamics of jet formation, enabling the study of processes influenced by the type of initiating parton and the transition from perturbative splittings to the confinement of partons into hadrons.

\begin{figure}[htb]
\centerline{%
\includegraphics[width=0.4\textwidth]{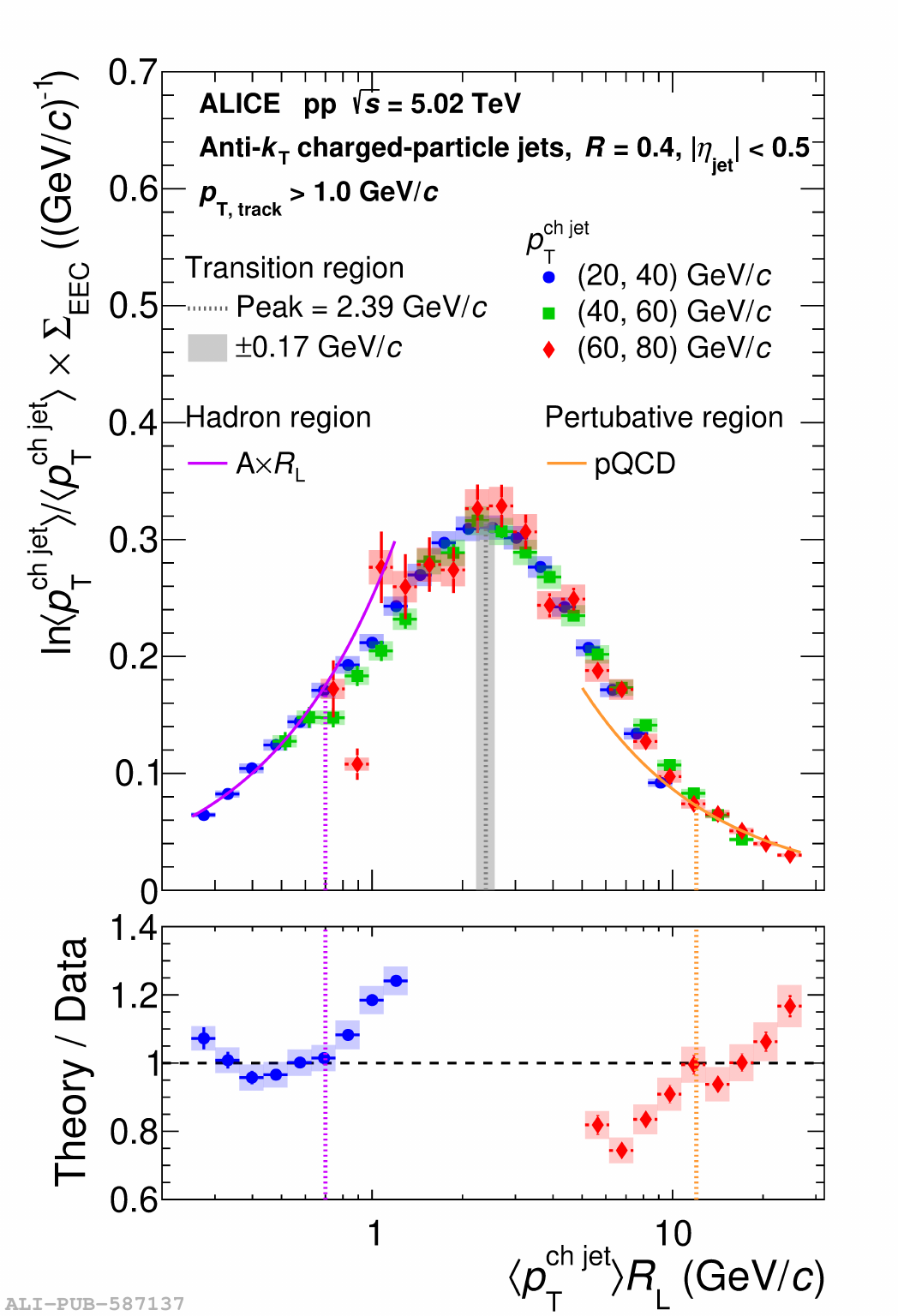}}
\caption{Normalized $\Sigma_{\text{EEC}}$ distributions as a function of $\langle\pTjch\rangle\RL$~\cite{ALICE:2024dfl}.}
\label{Fig:EEC}
\end{figure}

Figure~\ref{Fig:EEC} shows the EEC distributions measured in \pp collisions at $\sqrt{s}=5.02$ TeV for different $\pTjch$ intervals scaled by $\langle\pTjch\rangle\ln{\langle\pTjch\rangle}$ as a function of $\langle\pTjch\rangle\RL$, where $\langle\pTjch\rangle$ is the average \pT for the charged-particle jets in each \pT interval~\cite{ALICE:2024dfl}.
At large $\RL$, the jet structure is dominated by the perturbative splitting of partons during the initial stages of its evolution.
This region is compared to a next-leading-logarithmic pQCD calculation~\cite{Lee:2022ige}.
The data deviates from the perturbative scaling as \RL decreases, which is expected as non-perturbative effects become important.
The small-\RL region corresponds to later times in the jet evolution where the partons evolve into hadrons, referred to as the non-perturbative or free hadron scaling region.
This region is well described by a linear function.
The linear scaling reflects a situation where energy is uniformly distributed within the jet, resulting in purely combinatorial correlations among the EEC pairs, characteristic of freely moving (non-interacting) hadrons.
The peak observed in the intermediate \RL region signifies the transition from the perturbative to the non-perturbative regime, reflecting the hadronization process where partons are confined into color-neutral hadrons.
This transition is marked by the turnover of the EEC peaks, occurring at $\langle\pTjch\rangle\RL = 2.39 \pm 0.17$~\GeVc across all  $\pTjch$ intervals.

\section{Inclusive jets in \pPb collisions}
The nuclear modification factor, \RpPbjch of charged-particle jet yield in minimum bias \pPb collisions is quantified by comparing the jet cross section in \pPb collisions normalized by the number of nucleons of the Pb ion, $A=208$, to the jet cross section in \pp collisions.

\begin{figure}[htb]
\centerline{%
\includegraphics[width=12.5cm]{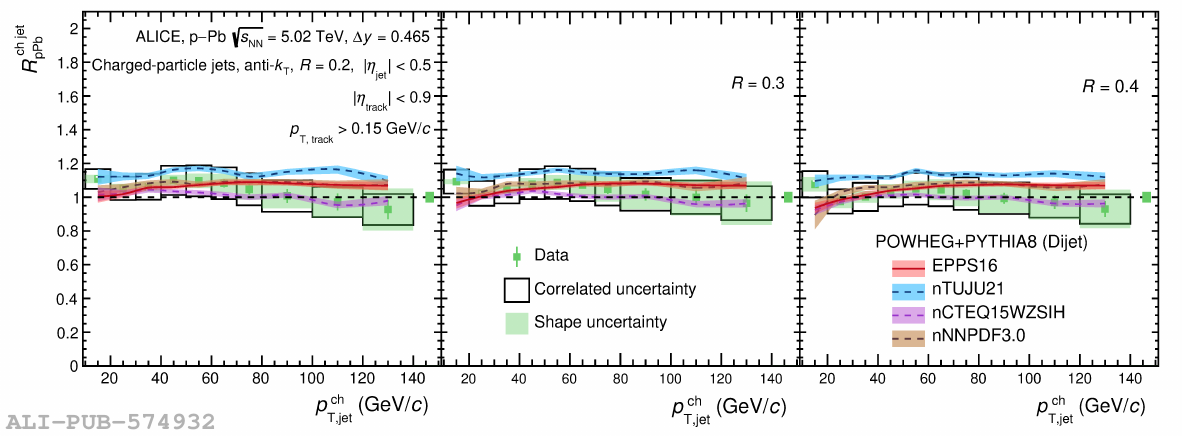}}
\caption{Inclusive charged-particle jet nuclear modification factor, \RpPbjch in \pPb collisions at \fivenn~\cite{ALICE:2023ama}.}
\label{Fig:RpPb}
\end{figure}

Figure~\ref{Fig:RpPb} depicts the \RpPbjch for jets with jet resolution parameters $R=0.2$, $0.3$, and $0.4$ as a function of \pTjch~\cite{ALICE:2023ama}.
The $\RpPbjch$ is compatible with unity within uncertainties in the reported transverse momentum range $10<\pTjch<140$~\GeVc and is approximately independent of jet transverse momentum and jet resolution parameter.
The results indicate that jet quenching, if present, is below the sensitivity of the current measurement.
The data are compared to the next-to-leading-order (NLO) prediction simulated using POWHEG+PYTHIA 8~\cite{Alioli:2010xa, Sjostrand:2014zea}.
The NLO prediction calculated using various nuclear-modified parton distribution functions (nPDFs) is in agreement with the data within uncertainties.
This shows that the effects of nPDFs have a minor impact on jet production.

\section{Semi-inclusive jets in \PbPb collisions}
This analysis utilizes a differential observable based on the semi-inclusive distribution of charged-particle jets recoiling from a high-\pTtrig hadron, as a function of the recoil jet transverse momentum \pTjch and the trigger-recoil jet azimuthal separation $\Delta\varphi$~\cite{ALICE:2023jye}.
The \Drecoil observable is the difference between the trigger-normalised yield of charged-particle jets in two exclusive trigger track intervals, \TTSig (TT\{20, 50\}) and \TTRef (TT\{5, 7\}),
{\tiny
\begin{equation}
    \DrecoilpTphi =
    \frac{1}{\Ntrig}\dNjetdpTdphi\Bigg\vert_{\pTtrig\in{\TTSig}}  - \cRef\times \frac{1}{\Ntrig}\dNjetdpTdphi\Bigg\vert_{\pTtrig\in{\TTRef}},
    \label{eq:DRecoil}
\end{equation}
}

\noindent
where \cRef is a normalization factor to account for conservation of jet density.
By analyzing the azimuthal angular separation between trigger hadrons and associated recoiling jets, these correlations provide insight into large-angle jet deflection in the QGP and transverse broadening effects.
Since the trigger-normalized jet yield is independent of \pTtrig, the uncorrelated background will be subtracted from \Drecoil, enabling jet measurements at lower jet \pT and large $R$~\cite{ALICE:2023jye, ALICE:2023qve}.

\begin{figure}[htb]
\centerline{%
\includegraphics[width=0.6\textwidth]{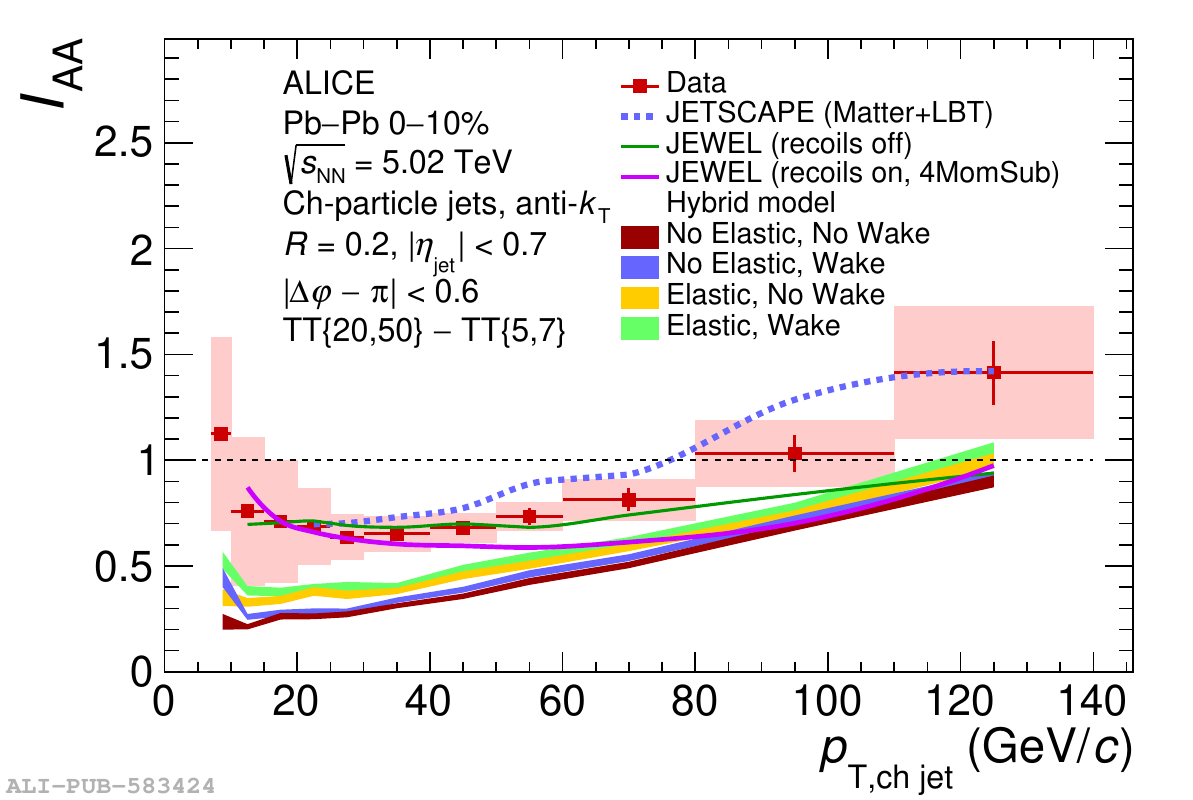}}
\caption{The $\IAA$ for recoil jets triggered by hadrons as a function of $\pTjch$ in \PbPb collisions at \fivenn compared to MC model simulations~\cite{ALICE:2023jye}.}
\label{Fig:IAA}
\end{figure}

The modification of yields due to medium effects in $\PbPb$ collisions is quantified by the $\IAA = \Delta_{\mathrm{recoil}}(\PbPb)/\Delta_{\mathrm{recoil}}(\pp)$.
Figure~\ref{Fig:IAA} shows that the $\IAA$ significantly depends on $\pTjch$~\cite{ALICE:2023jye}.
The $\IAA$ is above untiy for $\pTjch < 10$~\GeVc, and falls below unity for $10 < \pTjch < 80$~\GeVc, indicating medium-induced yield suppression due to energy loss.
Importantly, $\IAA > 1$ does not imply the absence of energy loss in jets.
The steeply falling spectrum of trigger hadrons suggests that these hadrons are predominantly produced near the surface of the QGP, where they can escape with minimal energy loss.
This phenomenon is referred to as surface bias.  
Additionally, model calculations indicate that energy loss experienced by trigger-side jets can also contribute to the enhancement of $\IAA$~\cite{He:2024rcv}.

The data are compared with several model predictions.
Among these, JETSCAPE~\cite{JETSCAPE:2022jer} best captures both the magnitude and the $\pTjch$ dependence of $\IAA$.
The JEWEL calculations~\cite{Zapp:2013vla}, both with recoils off and recoils on, describe the $\IAA$ at low $\pTjch$ but fail to reproduce the $\pTjch$ dependence of the data, underpredicting it at higher $\pTjch$.
The Hybrid Model with all of its variants~\cite{Casalderrey-Solana:2014bpa}, underestimate the magnitude of $\IAA$ across all \pT.

\section{Conclusion}
We have presented new ALICE measurements of jets in \pp, \pPb, and \PbPb collisions.
These measurements provide valuable insights into the behavior of jets and their substructure.
The observables discussed not only highlight the current understanding of jet production and modification but also set the stage for future studies.
With the significant increase in data recorded by ALICE during Run $3$, it will be possible to achieve increasingly precise constraints on jet properties and their substructure. 
Furthermore, these high-precision measurements will offer enhanced opportunities to explore and understand the properties of the QGP, shedding light on its role on jet quenching and the dynamics of strongly interacting matter under extreme conditions.

\vspace{2cm}
\noindent
\textbf{Acknowledgment}
This work is supported by the National Key R \& D Program of China (Grant No. 2018YFE0104700, 2022YFE0116900, and 2022YFA1602103), National Natural Science Foundation of China (Grant No. 12175085 and 12061141008), and Fundamental Research Funds for the Central Universities (No. CCNU220N003).
\bibliographystyle{utphys}   
\bibliography{ActaPhysicaPolonicaB}
\end{document}